\documentstyle[aasms4]{article}

\renewcommand{\tabcolsep}{3pt}

\lefthead{Kikuchi et al.}
\righthead{Hot Region in the Virgo Cluster}

\begin{document}

\title{Detection of an X-Ray Hot Region in the Virgo Cluster of
Galaxies with {\it ASCA}}

\author{K. Kikuchi\altaffilmark{1}, 
C. Itoh\altaffilmark{2}, 
A. Kushino\altaffilmark{2}, 
T. Furusho\altaffilmark{2}, 
K. Matsushita\altaffilmark{2}, 
N. Y. Yamasaki\altaffilmark{2}, 
T. Ohashi\altaffilmark{2},
Y. Fukazawa\altaffilmark{3},
Y. Ikebe\altaffilmark{4},
H. B\"{o}hringer\altaffilmark{4}, and
H. Matsumoto\altaffilmark{5}}

\altaffiltext{1}{Space Utilization Research Program,
        National Space Development Agency of Japan,
        2-1-1 Sengen, Tsukuba, Ibaraki 305-8505, Japan; 
        Kikuchi.Kenichi@nasda.go.jp}
\altaffiltext{2}{Department of Physics, Tokyo Metropolitan University,
        1-1 Minamiosawa, Hachioji, Tokyo 192-0397, Japan}
\altaffiltext{3}{Department of Physics, University of Tokyo,
        7-3-1 Hongo, Bunkyo-ku, Tokyo 113-0033, Japan}
\altaffiltext{4}{Max-Planck-Institut f\"{u}r extraterrestrische Physik,
        Postfach 1603, D-85740, Garching, Germany}
\altaffiltext{5}{Center for Space Research NE80-6045,
        Massachusetts Institute of Technology,
        77 Massachusetts Avenue, Cambridge, MA02139-4307, USA}

\begin{abstract}

  Based on mapping observations with {\it ASCA}, an unusual hot region
  with a spatial extent of 1 square degree was discovered between M87
  and M49 at a center coordinate of R. A. = 12h 27m 36s and Dec.\ =
  $9^\circ18'$ (J2000)\@. The X-ray emission from the region has a
  $2-10$ keV flux of $1 \times 10^{-11}$ ergs s$^{-1}$ cm$^{-2}$ and a
  temperature of $kT \gtrsim 4$ keV, which is significantly higher
  than that in the surrounding medium of $\sim 2$ keV\@.  The internal
  thermal energy in the hot region is estimated to be $V n k T \sim
  10^{60}$ ergs with a gas density of $\sim 10^{-4}$ cm$^{-3}$.  A
  power-law spectrum with a photon index $1.7-2.3$ is also allowed by
  the data. The hot region suggests there is an energy input due to a
  shock which is probably caused by the motion of the gas associated
  with M49, infalling toward the M87 cluster with a velocity $\gtrsim
  1000$ km s$^{-1}$.

\end{abstract}


\section{Introduction}
                    
Recent X-ray observations are revealing significant large-scale
variations of temperature and surface brightness in many clusters,
providing evidence that clusters are evolving.
Hydrodynamic simulations show that clusters recently formed through
mergers should indicate a complex temperature structure, and become
more regular with time (e.g.\ Roettiger, Burns, \& Loken 1993;
Takizawa 1999).  Thus, spatial distributions of the temperature of the
intracluster medium (ICM) provide important clues about the dynamical
evolution and the present state of the cluster.

In this letter, we perform a detailed investigation on the temperature
structure of the ICM in the Virgo cluster, based on extensive mapping
observations with {\it ASCA} (Tanaka, Inoue, \& Holt 1994). This
nearest rich cluster enables {\it ASCA} to perform spatially resolved
spectroscopy with moderate spatial resolution, and hot-gas properties
can be studied in both galaxy scales ($< 100$ kpc) and in the whole
cluster scale ($> 1$ Mpc).  The Virgo cluster is thought to be a
dynamically young system as recognized from its irregular structure in
the optical and X-ray bands. Thus, the present mapping study of the
cluster should provide us with valuable information to investigate the
on-going heating process in the ICM\@.

We assume the distance to the Virgo cluster to be 20 Mpc (e.g.\ 
Federspiel et al.\ 1998), hence $1'$ angular separation at the cluster
corresponds to 5.8 kpc.  The solar number abundance of Fe relative
to H is taken as $4.68 \times 10^{-5}$ (Anders \& Grevesse 1989)
throughout this letter.

\section{Observation and Analysis}

\subsection{Observation}

The mapping observations of the Virgo cluster have been carried out in
December 1996 to December 1998, with 28 pointings and a total exposure
time of $\sim 500$ ksec (Matsumoto et al.\ 1999; Ohashi et al.\ 1999;
Yamasaki et al.\ 1999).  Together with the data in the archive, the
area covered with ASCA in the Virgo cluster is $\sim 10$ deg$^2$.
Figure 1 shows the ASCA observed regions overlayed on the X-ray
contours with {\it ROSAT} (B\"{o}hringer et al.\ 1994).  The radius of
the circles is $22'$ corresponding to GIS field of view (Makishima et
al.\ 1996; Ohashi et al.\ 1996).

We selected the GIS data observed with the telescope elevation angle
from the Earth rim $> 5^\circ$, and the data taken with unstable
attitude after maneuvers were discarded.  Flare-like events due to the
background fluctuation were also excluded (Ishisaki 1996).  The cosmic
X-ray background (CXB) was estimated from the archival data taken
during 1993--1994 (Ikebe 1995), and the long-term variability of the
non X-ray background of the GIS (Ishisaki 1996) was corrected for.

\subsection{Image Analysis}

To derive the pure ICM component, contaminating X-ray sources have to
be excluded.  We carried out a source detection analysis developed for
the CXB study by Ueda (Ueda et al.\ 1999), who dealt with the
complicated detector response in a systematic way including the
position- and energy-dependence of the point spread function (PSF) of
the {\it ASCA} X-ray telescope.  Pointings containing bright sources
such as M87, M49, A1553, and A1541 were excluded from the analysis,
leaving 29 pointings to be analyzed (indicated by blue circles in
Figure 1).  We adopt a rather low flux level in detecting the source
candidates, $3 \sigma$ above the background in $0.7-7$ keV band, since
our interest is in the remaining diffuse component.  The analysis
detected 231 source candidates with X-ray flux $\gtrsim 1 \times
10^{-13}$ erg cm$^{-2}$ s$^{-1}$ in the $2-10$ keV band and all of them
have been masked out from the mapping data (see Figure 1). The mask
regions centered on the candidate positions have radii depending on
the source flux, since brighter sources affect wider regions due to
the image spread by the PSF effect. The mask radius is determined
where the surface brightness due to the source drops to less than 5\%
of the ICM level.

To examine spatially resolved spectral features, we need to know the
surface brightness distribution in the whole cluster to estimate the
amount of the stray light, which consists of photons generated outside
the field of view (e.g.\ Honda et al.\ 1996). Fortunately, the RASS
({\it ROSAT} All-Sky Survey, B\"{o}hringer et al.\ 1994; Voges et al.\
1996) data is available for this purpose, and we can produce the
template of the brightness profile.  In the RASS image, diffuse soft
X-ray emission, which possibly comes from the rim of LOOP I (e.g.\
Raymond 1984; Egger \& Aschenbach 1995), is present in the direction
of the Virgo cluster (Snowden et al.\ 1995).  Therefore, we use the
PSPC data only above 0.9 keV to exclude possible soft X-ray
contamination.  Spatially uniform background, obtained from a blank
sky region, was subtracted from the Virgo RASS data.  Based on this
template image, a ray-tracing simulation (Tsusaka et al.\ 1995) for
the GIS observation was carried out assuming a uniform temperature of
2.0 keV and a metallicity of 0.2 solar, which are the typical
parameters in the Virgo field (e.g.\ Koyama, Takano, \& Tawara 1991;
Matsumoto et al.\ 1999).

As a result, we found that the intermediate region between M87 and M49
was almost free from stray light from M87 and M49\@. The
contaminating flux from these 2 galaxies is less than a few \%. 
Considering the complex spectral structures (such as temperature
and abundance gradients) in these galaxies, this makes the data
analysis for the intermediate region much easier.

\subsection{Spectral Analysis}

To derive spectral parameters (such as temperature and surface
brightness) in a region, we have to know parameters in the surrounding
regions to evaluate the contamination of stray light.  We carried out
a first-order estimation of the spectral parameters, adopting an
analysis method developed by Honda et al.\ (1996).  This is performed
by fitting individual spectra with a modified response functions that
partly compensates the effect of the stray light.  The RASS image
obtained in the previous section is used to estimate the amount of the
stray light.

A spectral analysis has been performed for each pointed region in
the $0.7-8$ keV band with a Raymond-Smith model (Raymond \& Smith 1977,
hereafter R-S model). The interstellar absorption $N_{\rm H}$ is fixed
to the Galactic value ($1.7-2.5 \times 10^{20}$ cm$^{-2}$).  The
temperature distribution derived from the GIS spectral fits is shown
in Figure 2, with a color-coded plot of temperature in the left panel
and a plot as a function of distance from M87 in the right panel,
respectively.

As shown in Figure 2, the temperature around M87 is $\sim 2.5$ keV and
slightly decreases to $\sim 2.0$ keV at $\sim 1^{\circ}$ away from M87
in the northwest region.  In the south region, the average temperature
at a distance of $2^{\circ}$ from M87 is still $\sim 2.5$ keV, with a
large scatter from 1.8 keV to 3.4 keV\@.  The metal abundance is
poorly constrained in most of the regions because of low photon
statistics.  We only mention that the best-fit values suggest that the
metal abundance in the general cluster regions is around 0.2 solar
with a scatter of about $\pm 0.2$ solar from position to position.  If
we fitted the spectra with free absorption, the data generally require
that no absorption (even the Galactic $N_{\rm H}$) is present.  This
suggests existence of an additional soft component below $\sim 1$ keV,
which may be the foreground emission of the Galactic soft X-rays.

\section{The Hot Region}

As shown in Figure 2, three regions W1, W2 and W3 along the ``emission
bridge'' between M87 and M49 show the ICM temperatures rising to $\sim
3$ keV\@.  To improve the statistics, the three spectra are combined
(hereafter called W123) because individual fits indicate statistically
the same temperature.  For comparison, the spectra for regions E1, E2,
and E3 (hereafter E123), just to the east of W123, are also combined.
Both W123 and E123 regions are elongated in parallel to the ``emission
bridge'', and the distance from M87 and M49 is almost the same.
Errors in the contaminating spectra from nearby bright sources
(i.e. A1553, NGC 4325, A1541, and QSO 1225+089) and those in the
remaining fluxes of masked-out sources are the major origin of the
systematic error for temperatures in W123 and E123. This error is
found to be less than 0.2 keV, and its effect works on the two
temperatures in the same sense: thus keeping the temperature
difference the same.

The single-temperature model gives a poor fit in the energy range
$0.7-8.0$ keV for the 2 spectra W123 and E123\@. As shown in Table 1,
the best-fit results are $\chi^2/\nu = 45.2/21$ and $31.1/21$ for W123
and E123, respectively.  This is mainly due to excess emission below 1
keV in both spectra.  A two-component (R-S and a soft thermal
bremsstrahlung) model improves the fit to $\chi^2/\nu = 21.8/20$ and
$\chi^2/\nu = 27.2/20$ for W123 and E123, respectively (see Figure 3).
The additional thermal bremsstrahlung component yields the best-fit
temperature $kT = 0.2-0.3$ keV with $F_{\rm X} \sim 0.8 \times
10^{-15}$ erg cm$^{-2}$ s$^{-1}$ arcmin$^{-2}$ for $0.5-2$ keV for both
W123 and E123 spectra.  These values are close to the ROSAT result
($kT = 0.15$ keV and $F_{\rm X} = 0.6 \times 10^{-15}$ ergs cm$^{-2}$
s$^{-1}$ arcmin$^{-2}$ in $0.5-2$ keV band, Irwin \& Sarazin (1996)),
supporting the view that the soft emission is due to the Galactic hot
interstellar medium.  These results indicate that the region W123 has
a significantly high temperature of $\sim 4$ keV, while E123, just in
the east of W123, shows $\sim 2$ keV which is the typical temperature
of the Virgo cluster.

Based on these results, we neglect the energy range below 2 keV and
look into the pure ICM component in the region W123\@.  We also
subtracted contaminating photons, which come from the surrounding
region of W123, assuming the temperature and metallicity of the
surrounding ICM are 2 keV and 0.2 solar, respectively. Fixing the
abundance to 0.2 solar, acceptable fits with R-S model are obtained
with $\chi^2/\nu = 10.5/11$ (Table 1).  The ray-tracing simulation
gives a systematic error for the stray-light intensity by $-30\% -
+10\%$ as estimated from offset observations of the Crab nebula
(Ishisaki 1996). The systematic error due to a fluctuation of the CXB
flux is $\sim 10$\% for the GIS field of view, and the uncertainty in
the estimation of the non X-ray background level is $6$\%.  Including
all these errors, we can conclude that the temperature in W123 is
still higher than that in E123 with more than 90\% confidence.  The
total flux of the hot region in $2-10$ keV band is ($9.2-10.3$)
$\times 10^{-12}$ erg cm$^{-2}$ s$^{-1}$ at the 90\% confidence limit,
and the emission measure $\int n^2 dl$ is estimated to be ($3.4-5.0$)
$\times 10^{16}$ cm$^{-5}$ assuming an extent of the hot region to be
$4 \times 10^3$ arcmin$^2$.

So far, the ``hot'' emission has been assumed to have a thermal
spectrum.  However, the data also allow non-thermal (power-law)
models. Spectral fit for the $2-8$ keV hot region data (W123) with
a power-law model gives an acceptable result of $\chi^2/\nu = 9.9/11$
with a photon index between $1.7-2.3$ at the 90\% confidence (see
Table 1). Since no significant Fe-K line is seen in the W123 spectrum
with $EW \le 821$ eV for a 6.7 keV line at the 90\% confidence, the
diffuse non-thermal emission remains as a possibility from the ASCA
observations.

\section{Discussion}

The previous {\it Ginga} observations have suggested a temperature
rise in the ICM from M87 to M49 (Takano 1990; Koyama, Takano, \&
Tawara 1991). However, {\it ROSAT} data showed no such evidence
 (B\"{o}hringer et al.\ 1994) and implied a possibility that the
non-imaging {\it Ginga} data were contaminated by background sources.
The extensive mapping observations from {\it ASCA} have shown the
correct temperature structure in the Virgo cluster for the first time
and unambiguously detected an unusual ``hot'' region in the
ICM\@. This detection provides a clear evidence that the Virgo cluster
is a young system in which a local gas heating is taking place now in
the cluster outskirts.

The emission measure of the hot region W123 obtained in the previous
section gives a rough estimate of the gas density $n$ to be of the
order of $1 \times 10^{-4}$ cm$^{-3}$. Here, we assume that the
line-of-sight depth of the hot component is $\sim 300$ kpc which is
the same order as the projected length of the region. The internal
thermal energy of the hot component, $E_{\rm th} = V n kT$ where $V$
and $T$ are volume and temperature, is calculated as $\sim 10^{60}$
ergs.  This level of energy is orders of magnitude lower than the
kinetic energy involved in a typical subcluster merger ($10^{63-64}$
ergs).  This means that if there is a bulk motion of gas in this
region with $v \approx 1000$ km s$^{-1}$, caused by an infall of
galaxies or a small group of galaxies, then it can supply enough
energy to heat up the gas to the observed temperature.

One might feel some difficulty to heat up a localized spot by a
merger, however, such a local effect could be produced by the
merging of a subclump of the irregular M49 subcluster.
Time scale for thermal conduction is roughly estimated as 
$t_{\rm cond} \approx 8 \times 10^8$ yr for the gas density in the hot
region and the scale length of the temperature gradient to be 500 kpc.
If the gas moving with $v \approx 1000$ km s$^{-1}$ receives some heat
input, the hot region would become elongated by $\sim 800$ kpc because
of the slow heat conduction. This situation could be related with the
observed north-south elongation of the hot region.

Honda et al.\ (1996) reported temperature variation of the ICM in the
Coma cluster, and found a remarkable hot region ($\gtrsim 11$ keV)
which is distinct from the average temperature of the whole cluster
($\sim 8$ keV).  This hot region is located at $40'$ (1.6 Mpc) offset
from the cluster center, and has an angular extent of $\sim 20'$
radius. We can roughly estimate the extra internal energy in the Coma
hot region as $\sim 8 \times 10^{61}$ ergs, which is nearly 2 orders of
magnitude higher than that in the Virgo case.

Irwin \& Sarazin (1996) discuss that M49 is moving supersonically ($v
\approx 1300$ km s$^{-1}$) in the Virgo ICM toward the direction of
M87\@.  The gravitational mass of M49 subcluster is estimated as $8.7
\times 10^{13} M_{\odot}$ from the {\it ROSAT} observation (Schindler,
Binggeli, \& B\"{o}hringer 1999).  Then the kinetic energy of the M49
subcluster is roughly estimated as $1 \times 10^{63}$ ergs, which is
sufficiently large to heat up the hot-region gas.  Using
Rankine-Hugoniot jump condition (e.g.\ Shu 1992), the Mach number of
the shock wave to heat up 2 keV gas to $\sim 4$ keV should be $\sim
2$. Since the sound velocity of the 2 keV gas is $\sim 700$ km
s$^{-1}$, the required velocity is close to the Irwin \& Sarazin
result.

Hard X-ray ($kT \gtrsim 10$ keV) emission from clusters of galaxies
was reported from previous observations (e.g.\ Fusco-Femiano et al.\
1999 for Coma cluster), and the existence of non-thermal emission has
been suggested.  For the hot region detected here, we cannot confirm
whether the origin of the hard emission is thermal or non-thermal,
because of the lack of observational evidences.  If we assume that
relativistic electrons are produced by first-order Fermi acceleration,
the momentum spectrum of the electrons is described as $N(p) = N_0
p^{- \mu}$. Here, $\mu = (r + 2) / (r - 1)$ and $r$ is a ratio of the
shock compression.  For the shock with a Mach number $\sim 2$, the
exponent is implied as $\mu \approx 3.3$.  In such a steep spectra,
an energy loss due to nonthermal bremsstrahlung dominates the inverse
Compton loss (Sarazin \& Kempner 1999).  The electrons lose their
energy through Coulomb loss, whose time scale is estimated as $t_{\rm
Coul} \approx 3 \times 10^8 \gamma$ yr. This is similar to the $t_{\rm
cond}$ estimated above.

Above considerations suggest that in both thermal and non-thermal
cases, the extra energy built up in the hot region would dissipate
away within about 1 Gyr, due to thermal conduction or Coulomb
loss. Therefore, it seems likely that the energy supply into the
hot region has started only within the past 1 Gyr, or
alternatively a long continuous supply of energy has been occurring
here over a cosmological time scale. The infall of the M49 subcluster
can supply energy into ICM for a very long time and is probably
connected with the local gas heating as detected in the Virgo cluster.

\acknowledgments We thank Y. Ueda and Y. Ishisaki for their support of
source detection analysis and background estimation. Stimulating
discussion with T. Reiprich, C. Sarazin, K. Masai, S. Okamura and
M. Takizawa are also acknowledged. K. K. acknowledges hospitality in
MPE and support from the Japan Science and Technology Corporation
(JST)\@.  This work is partly supported by the Grants-in Aid of the
Ministry of Education, Science, Sports and Culture of Japan, 08404010.

\begin{table*}[htb]
  \begin{center} 
   \caption{The best-fit spectral parameters of the linking region between 
            M87 and M49}
  \begin{tabular*}{\textwidth}{@{\hspace{\tabcolsep}
    \extracolsep{\fill}}lllllll}    \hline\hline
    \multicolumn{1}{l}{Region} 
  & \multicolumn{2}{l}{Model} 
  & \multicolumn{3}{l}{Parameters}
  & \multicolumn{1}{l}{$\chi^2/\nu$} \\
    \cline{1-1} \cline{2-3} \cline{4-6} \cline{7-7}
    \hline
    E123 & single component$^a$ 
         & R-S 
         & $kT = 2.10_{-0.25}^{+0.27}$,
         & $Z = 0.29_{-0.20}^{+0.32}$,
         & $F_{\rm X,2-10} = 1.21_{-0.15}^{+0.16}$ 
         & 31.1/21 \\
    \cline{2-7}
         & two components$^a$
         & R-S 
         & $kT = 2.25_{-0.29}^{+0.59}$,
         & $Z = 0.2$ (fixed), 
         & $F_{\rm X,2-10} = 1.26_{-0.15}^{+0.12}$ 
         & 27.2/20 \\
         & 
         & Brems
         & \multicolumn{2}{l}{$kT = 0.20$ (unconstrained)}, 
         & $F_{\rm X,0.5-2} = 0.58_{-0.41}^{+1.67}$ 
         & \\
    \hline
    W123 & single component$^a$ 
         & R-S 
         & $kT = 3.14_{-0.34}^{+0.31}$,
         & $Z = 0.63_{-0.30}^{+0.39}$,
         & $F_{\rm X,2-10} = 2.04_{-0.21}^{+0.19}$ 
         & 45.2/21 \\
    \cline{2-7}
         & two components$^a$
         & R-S 
         & $kT = 4.31_{-0.81}^{+1.11}$,
         & $Z = 0.2$ (fixed), 
         & $F_{\rm X,2-10} = 2.22_{-0.20}^{+0.17}$ 
         & 21.8/20 \\
         & 
         & Brems
         & $kT = 0.32_{-0.11}^{+0.22}$,
         & 
         & $F_{\rm X,0.5-2} = 0.95_{-0.12}^{+0.27}$
         & \\
    \cline{2-7}
         & single component$^b$
         & R-S 
         & $kT = 5.32_{-1.52}^{+3.70}$,
         & $Z = 0.2$ (fixed), 
         & $F_{\rm X,2-10} = 2.45_{-0.28}^{+0.30}$ 
         & 10.5/11 \\
    \cline{2-7}
         & single component$^b$
         & Power-law
         & $\Gamma = 1.97_{-0.27}^{+0.28}$,
         & 
         & $F_{\rm X,2-10} = 2.59_{-0.26}^{+0.26}$
         & 9.9/11 \\
    \hline
    \hline 
  \end{tabular*}
\end{center}
\tablecomments{ The $kT$ represents temperature in keV, $Z$ is heavy
  element abundance in solar unit, and $\Gamma$ is photon index.  The 
  flux $F_{\rm X,2-10}$ and $F_{\rm X,0.5-2}$ are in $10^{-15}$ ergs
  sec$^{-1}$ cm$^{-2}$ arcmin$^{-2}$ in $2-10$ and $0.5-2$ keV range,
  respectively.  The errors in all parameters represent 90\%
  confidence limits.  The R-S and Power-law components were modified
  by interstellar absorption with $N_{\rm H} = 1.9 \times 10^{20}$
  cm$^{-2}$.}
  $^a$ The data in the energy range of $0.7-8$ keV were used.\\
  $^b$ Only the data in the energy range $2-8$ keV were used, and
   the contaminating photons outside of the region are subtracted 
   assuming the temperature of 2 keV and metal abundance of 0.2 solar.\\
\end{table*}

\clearpage 
\figcaption
  {{\it ASCA} observed regions superposed on the {\it ROSAT} PSPC
  contour in $0.5-2$ keV band.  Contours show the X-Ray intensity
  observed by the {\it ROSAT} All-Sky Survey in a logarithmic scale,
  increasing by factors of 1.2.  Red and blue circles with radii $22'$
  indicate the GIS observed fields, and the red regions all containing
  bright sources are excluded in the present analysis.  The mask
  regions to exclude contaminating sources are also indicated by
  filled green circles.
  \label{fig1}}

\figcaption
  {Temperature distribution of the Virgo cluster.
  The spectra of $0.7-8$ keV band were fitted with a R-S model.
  {\it left:} Temperature scale for the middle panel.
  {\it middle:} Two dimensional plot of the temperature distribution
  in the Virgo region.
  {\it right:} The temperature distribution as a function of distance from
  M87. The error bars indicate statistical errors. Triangles show
  temperatures at W1, W2, and W3, and rectangles are at E1, E2, and E3.
  \label{fig2}}

\figcaption
  {GIS spectra of W123 (red) and E123
  (blue). Spectral data are shown with crosses, and the best-fit
  spectra of the two-component (R-S model and thermal
  bremsstrahlung) model are shown with solid lines. The best-fit
  thermal bremsstrahlung components are also indicated with dashed
  lines. See table 1 for the error ranges of the spectral
  parameters.
  \label{fig3}}

\end{document}